\documentclass[aps,12pt,superscriptaddress,amsfonts,amssymb,amsmath]{revtex4}
\pdfoutput=1

\usepackage{graphicx}
\usepackage{epsfig}
\usepackage{makeidx}
\usepackage{epstopdf}
\usepackage{natbib}
\usepackage{xcolor}
\usepackage{subcaption}
\usepackage{amsmath}
\usepackage{appendix}

\begin{document}

\title{Aharonov-Bohm electrodynamics in 
material media: a scalar e.m.\ field cannot cause dissipation in a 
medium}

\author{F.\ Minotti \footnote{Email address: minotti@df.uba.ar}}
\affiliation{Universidad de Buenos Aires, Facultad de Ciencias Exactas y Naturales, Departamento de F\'{\i}sica, Buenos Aires, Argentina}
\affiliation{CONICET-Universidad de Buenos Aires, Instituto de F\'{\i}sica Interdisciplinaria y Aplicada (INFINA), Buenos Aires, Argentina}

\author{G.\ Modanese \footnote{Email address: giovanni.modanese@unibz.it}}
\affiliation{Free University of Bozen-Bolzano \\ Faculty of Science and Technology \\ I-39100 Bolzano, Italy}

\linespread{0.9}

\begin{abstract}

In the extension of Maxwell equations based on the Aharonov-Bohm Lagrangian the e.m.\ field has an additional degree of freedom, namely a scalar field generated by charge and currents that are not locally conserved. We analyze the propagation of this scalar field through two different media (a pure dielectric and an ohmic conductor) in a range of frequencies such that the properties of the media are independent from the frequency. We find that an e.m.\ scalar wave cannot propagate in a material
medium. If a scalar wave in vacuum impinges on a material medium it is reflected, at most exciting in the medium a pure ``potential'' wave (which
we also call a ``gauge'' wave) propagating at $c$, the speed of light in vacuum, with a
vector potential whose Fourier amplitude is related to that of the scalar
potential by $\omega \mathbf{A}_{0}=\mathbf{k}\phi _{0}$, where $\omega^{2}=c^{2}\left\vert \mathbf{k}\right\vert^{2}$.

\end{abstract}

\maketitle

\section{Introduction}

The extension of Maxwell equations according to the Lagrangian proposed by Aharonov and Bohm has been widely studied in the last years, both theoretically and from the point of view of possible applications. This extension is quite natural and was actually introduced, with various motivations and technical approaches, even before the work by Aharonov and Bohm \cite{ohmura1956new, aharonov1963further, alicki1978generalised,cornille1990propagation,van2001generalisation, woodside2009three,  hively2012toward,arbab2017extended}. For recent reviews and developments see \cite{Modanese2017MPLB,modanese2017electromagnetic,modanese2019design,modanese2019high,hively2019classical,keller2019electrodynamics,keller2019ohmura,reed2020implications,minotti2021quantum,minotti2021current,Hively_2021,minotti2022electromagnetic,cornille2022critical}.

The extended theory is also called ``scalar electrodynamics'', because a crucial role is played in it by a scalar field which is not present in Maxwell theory and represents in fact a further degree of freedom of the e.m.\ field. The uniqueness of the extended theory, under the usual assumptions of regularity and relativistic invariance, has been proven by Woodside \cite{woodside2009three}. The Aharonov-Bohm action is characterized by a reduced gauge invariance (see  \cite{Modanese2017MPLB}) and has been investigated until now only at the classical level, except (under some restrictive assumptions) in the work by Jimenez and Maroto \cite{jimenez2011cosmological}. Energy and momentum conservation has been studied in detail in \cite{minotti2021quantum}.

The scalar field, which in this work will be denoted by $S$, can be expressed in terms of the e.m.\ potentials as $S=\mu_0 \varepsilon_0 \partial_t \phi+\nabla \cdot \mathbf{A}$. Due to the reduced gauge invariance, $S$ cannot be set identically to zero like in Maxwell theory, but its occurrence is rare because its source is the ``extra-current'' $I=\partial_t \rho+\nabla \cdot \mathbf{J}$, a quantity which is different from zero only where charge is not locally conserved.

For classical sources, within a particle-like description of charge and current where it is always possible to ``count'' the particles crossing a given surface in a given time, violations of local charge conservation are inconceivable. For quantum sources, however, the concept of exact localization of particles is replaced by the concept of probability expressed through a wavefunction, and the expectation values of physical quantities are subjected to uncertainties, also in macroscopic wavefunctions like those of, e.g., tunnelling Josephson junctions \cite{minotti2021quantum}. 

Moreover, even though in the solutions of the Schr\"odinger equation with a local potential and in free quantum field theory the local conservation of probability is guaranteed, several physical systems are described by Schr\"odinger equations with non-local potentials \cite{lenzi2008solutions, lenzi2008fractional, caspi2000enhanced, chamon1997nonlocal, balantekin1998green, laskin2002fractional, wei2016comment,modanese2018time} or by renormalized quantum field theories where anomalies cannot be excluded \cite{cheng1984gauge,parameswaran2014probing,chernodub2022thermal}. Some first-principles numerical calculations of the currents in molecular devices have shown anomalies in local conservation, whose interpretation is still unclear \cite{li2008definition,zhang2011first,cabra2018simulation,jensen2019current,garner2019helical,garner2020three}. Certain anomalies might be dismissed as artefacts of the mathematical models, but then, how is it possible in quantum mechanics to speak of a reality beyond mathematical models?

The conclusion is, in our opinion, that the generation of an e.m.\ scalar field $S$ by quantum systems is rare but cannot be excluded. This process is described by the extended theory in vacuum in a relativistically invariant way and does not involve any violation of causality. A detailed calculation of the radiation field generated by anomalous sources oscillating at high frequency in extended electrodynamics has been performed in \cite{minotti2021current}.

The next big question then is: what happens when the scalar field encounters some medium and propagates in it? The answer is not simple, because the behavior of a scalar field is markedly different from the familiar behavior of electric and magnetic fields. Moreover, the presence of a scalar field usually implies (in vacuum) the presence of a longitudinal component in propagating electric fields. A simple-minded approach to this issue can lead to paradoxes, as we will show in the following (Sect.\ \ref{paradox}). For a full description one must take into account the propagation equations of the potentials $\phi$, $\mathbf{A}$, and the startling conclusion is the following (Sect.\ \ref{equ-pot}): the scalar field cannot enter a medium in which local conservation of charge holds, and correspondingly the longitudinal component of $\mathbf{E}$ also vanishes in the medium. This is true both for insulating dielectrics and for conductors in which $\mathbf{j}=\sigma \mathbf{E}$, at least in the range of frequencies where the conductivity $\sigma$ can be regarded as independent from the frequency. Only a ``pure potential'' wave propagates in the medium, with zero scalar field and zero longitudinal electric field. In Sect.\ \ref{reflection} we compute in detail reflection and transmission of all field components at a vacuum-material interface, obtaining the usual Fresnel relations plus the conditions for the potential waves. Finally, Sect.\ \ref{conclusions} contains our conclusions and outlook.

\section{Aharonov-Bohm electrodynamics in a material medium}

In Subsect.\ \ref{ext-vac} of this Section we recall the extended field equations in vacuum, in the presence of free charges. The non-Maxwellian scalar field $S$ which appears in the equations can only be generated by charge and currents which are not locally conserved (eq.\ \ref{Sequ}). Therefore we shall suppose that somewhere \emph{outside the medium} there exist such sources for $S$. On the other hand, we assume that \emph{in the medium} which the propagating fields encounter, local conservation does hold true, both for the free charges and currents $(\rho,\mathbf{j})$ and for the molecular charges and currents $(\rho_m,\mathbf{j}_m)$, related to polarization and magnetization in the medium. This assumption allows to reduce the complexity of the problem to an acceptable level. One could, more generally,  estimate the higher order corrections due to the possible failure of local conservation also inside the medium; such corrections turn out to be very small.

In Sect.\ \ref{ext-med} we write the extended field equations in a medium and their general solution in plane waves, which exhibits an apparent paradox concerning the longitudinal electric component $\mathbf{E}_{L0}$. The paradox is resolved in Sect.\ \ref{equ-pot}, where we solve the equations for the potentials $\phi$, $\mathbf{A}$, distinguishing between the case of a pure dielectric without losses and the case of an ohmic conductor. We show that in both cases one has in the medium $S=0$ and consequently $\mathbf{E}_{L0}=0$.

\subsection{Extended field equations in vacuum}
\label{ext-vac}

In this section we write the field equations of the extended Aharonov-Bohm electrodynamics in vacuum, in the presence of free charges and currents $(\rho,\mathbf{j})$ which do not satisfy everywhere the continuity relation $\partial_t \rho+\nabla \cdot \mathbf{J}=0$. These equations look like extended Maxwell equations in the sense that they contain a scalar field $S$ giving rise to an additional scalar source $\partial_t S$ for $\nabla \cdot \mathbf{E}$ in the first equation (\ref{vac1}) and to an additional vector source $\nabla S$ for $\nabla \times \mathbf{B}$ in the fourth equation (\ref{vac4}). From the second and third equation (\ref{vac2}) and (\ref{vac3}), which are unchanged compared to Maxwell's formulation, it follows that $\mathbf{E}$ and $\mathbf{B}$ can be expressed in terms of the potentials $\phi$, $\mathbf{A}$ in the usual way (\ref{vac5}) and (\ref{vac6}). Finally, one defines $S$ as $S=\mu_0 \varepsilon_0 \partial_t \phi+\nabla \cdot \mathbf{A}$ (eq.\ (\ref{vac7})). It is understood that the Lagrangian of the field is here the Aharonov-Bohm Lagrangian with reduced gauge invariance, namely (in covariant form and with Heaviside units, for brevity)
\begin{equation}
    L_{A.-B.}=-\frac{1}{4}F_{\mu}F^{\mu\nu}+\frac{1}{2}\lambda (\partial_\mu A^\nu)^2
\end{equation}
so that in general it is not possible to choose the gauge in such a way as to obtain $S=0$. The reduced gauge transformations are of the form
\begin{equation}
    A^\mu \to A^\mu+\partial^\mu \chi, \ \ \ \ \ \mathrm{with} \ \ \partial^\mu \partial_\mu \chi=0
\end{equation}

The extended equations valid in vacuum are, in summary,
\begin{subequations}
\begin{eqnarray}
\nabla \cdot \mathbf{E} &=&\frac{\rho }{\varepsilon _{0}}-\frac{\partial S}{
\partial t}, \label{vac1}\\
\nabla \cdot \mathbf{B} &=&0, \label{vac2}\\
\nabla \times \mathbf{E} &=&-\frac{\partial \mathbf{B}}{\partial t}, \label{vac3}\\
\nabla \times \mathbf{B} &=&\mu _{0}\varepsilon _{0}\frac{\partial \mathbf{E}
}{\partial t}+\mu _{0}\mathbf{j}+\nabla S, \label{vac4}\\
\mathbf{E} &=&-\nabla \phi -\frac{\partial \mathbf{A}}{\partial t}, \label{vac5}\\
\mathbf{B} &=&\nabla \times \mathbf{A}, \label{vac6}\\
S &=&\mu _{0}\varepsilon _{0}\frac{\partial \phi }{\partial t}+\nabla \cdot \mathbf{A}. \label{vac7}
\end{eqnarray}
\end{subequations}

\subsection{Extended field equations in a medium}
\label{ext-med}

Let us consider a medium with molecular charge density $\rho_m$ and molecular current density $\mathbf{j}_m$ connected to the magnetization field $\mathbf{M}$ and the polarization field $\mathbf{P}$ by the usual relations
\begin{equation}
    \rho _{m}=-\nabla \cdot \mathbf{P}, \ \ \ \ \ \mathbf{j}_{m}=\nabla \times \mathbf{M}+\partial_t \mathbf{P}\label{materialsources}
\end{equation}
These expressions satisfy by construction local conservation:
\begin{equation}
    \partial_t \rho _{m} +\nabla \cdot \mathbf{j}_{m}=0
\end{equation}

After including the sources (\ref{materialsources}) in Eqs. (\ref{vac1}) and (\ref{vac4}), and defining in the familiar way the auxiliary vectors $\mathbf{D}$ and $\mathbf{H}$ as
\begin{eqnarray}
\mathbf{D} &=&\varepsilon _{0}\mathbf{E}+\mathbf{P}, \\
\mathbf{H} &=&\mathbf{B}/\mu _{0}-\mathbf{M},
\end{eqnarray}
 we obtain the field equations
\begin{subequations}
\label{Eqmedgen}
\begin{eqnarray}
\nabla \cdot \mathbf{D} &=&\rho -\varepsilon _{0}\frac{\partial S}{\partial t%
},  \label{eqD} \\
\nabla \cdot \mathbf{B} &=&0,  \label{eqB} \\
\nabla \times \mathbf{E} &=&-\frac{\partial \mathbf{B}}{\partial t},
\label{eqE} \\
\nabla \times \mathbf{H} &=&\frac{\partial \mathbf{D}}{\partial t}+\mathbf{j}%
+\frac{1}{\mu _{0}}\nabla S.  \label{eqH}
\end{eqnarray}
plus the equation for $S$ (consistent with the
local conservation of the molecular sources): 
\end{subequations}
\begin{equation}
\mu _{0}\varepsilon _{0}\frac{\partial ^{2}S}{\partial t^{2}}-\nabla
^{2}S=\mu _{0}\left( \frac{\partial \rho }{\partial t}+\nabla \cdot \mathbf{j}\right) .  \label{Sequ}
\end{equation}

Using the usual constitutive relations for a homogeneous, isotropic medium
\begin{equation}
    \mathbf{D}=\varepsilon \mathbf{E}, \ \ \ \ \ \mathbf{B}=\mu \mathbf{H}
\end{equation}
one has the system 
\begin{subequations}
\label{ABmedia}
\begin{eqnarray}
\nabla \cdot \mathbf{E} &=&\frac{\rho }{\varepsilon }-\frac{\varepsilon _{0}%
}{\varepsilon }\frac{\partial S}{\partial t},  \label{gauss} \\
\nabla \cdot \mathbf{B} &=&0,  \label{nomonop} \\
\nabla \times \mathbf{E} &=&-\frac{\partial \mathbf{B}}{\partial t},
\label{faraday} \\
\nabla \times \mathbf{B} &=&\mu \varepsilon \frac{\partial \mathbf{E}}{%
\partial t}+\mu \mathbf{j}+\frac{\mu }{\mu _{0}}\nabla S.  \label{ampere}
\end{eqnarray}
\end{subequations}

\subsection{Wave solution for the electric field in a conductive medium and the dissipation paradox}
\label{paradox}

Now suppose further that
\begin{equation}
    \mathbf{j}=\sigma \mathbf{E}
\end{equation}
and that local conservation also holds, as previously discussed, for any free charges in the medium, namely
\begin{equation}
    \partial_t \rho +\nabla \cdot \mathbf{j}=0 \ \ \ \ \mathrm{(in \ the \ medium)}
\end{equation}

By eliminating the magnetic and scalar fields among these equations one obtains the relation 

\begin{equation}
\nabla ^{2}\mathbf{E}+\mu \varepsilon \omega ^{2}\mathbf{E}=-i\omega \mu
\sigma \mathbf{E}-\left( i\frac{\sigma }{\omega \varepsilon }\frac{\mu
\varepsilon }{\mu _{0}\varepsilon _{0}}+\frac{\mu \varepsilon }{\mu
_{0}\varepsilon _{0}}-1\right) \nabla \left( \nabla \cdot \mathbf{E}\right) .
\label{E_metal_right}
\end{equation}

The usual transverse mode $\mathbf{E}_{T}=\mathbf{E}_{T0}\exp \left( i%
\mathbf{k}\cdot \mathbf{x}\right) $, with $\mathbf{k}\cdot \mathbf{E}_{T0}=0$, satisfies 
\begin{eqnarray*}
\nabla \cdot \mathbf{E}_{T} &=&0, \\
\nabla \times \mathbf{E}_{T} &=&i\mathbf{k}\times \mathbf{E}_{T},
\end{eqnarray*}
so that the corresponding dispersion relation is 
\begin{equation}
k^{2}=\mu \varepsilon \omega ^{2}\left( 1+i\frac{\sigma }{\varepsilon \omega 
}\right) .  \label{k2trans}
\end{equation}
Note that in a good conductor at microwave frequencies $\frac{\sigma }{
\varepsilon \omega }\gg 1$, implying a strong dissipation of the
transverse mode. The consideration of microwave frequency here is only for convenience of possible experiments and in order to be in a regime dominated by the electrical conductivity (the opposite limit corresponding to a dielectric). Also at much larger frequencies, about the electron plasma frequency, the conductivity is strongly frequency dependent.

By replacing into Eq.\ (\ref{E_metal_right}) a longitudinal mode with Fourier amplitude $\mathbf{E}_{L0}$ proportional to $\mathbf{k}$, one obtains the relation
\begin{equation*}
\left( \omega ^{2}-\frac{k^{2}}{\mu _{0}\varepsilon _{0}}\right) \left( 1+i
\frac{\sigma }{\varepsilon \omega }\right) \mathbf{k}\cdot \mathbf{E}_{L0}=0,
\end{equation*}
from which, under the assumption that $\mathbf{k}\cdot \mathbf{E}_{L0}\neq 0$,
would result the dispersion relation $k^{2}=\mu _{0}\varepsilon _{0}\omega ^{2}$.
This result is rather surprising because it shows no decay, while a decay
should be present, since the power dissipated per unit volume is $\mathbf{j}\cdot 
\mathbf{E}=\sigma \left\vert \mathbf{E}\right\vert ^{2}$, which has a
positive time average. The answer to this paradox can be obtained considering the equations for the potentials (Sect.\ \ref{equ-pot}), which show that there is
actually no electric field (so that the assumption $\mathbf{k}\cdot \mathbf{E}_{L0}\neq 0$ is incorrect), and no scalar field either, in the longitudinal
mode in a material medium. It turns out that this mode is a ``pure potential'', or ``gauge'' wave, with zero fields, and zero power. The same happens also in a
dielectric without losses; by this we mean a material with real permittivity, or in other words the regime considered corresponds to the transparent region of dielectrics.

\subsection{Equations for the potentials and solution of the paradox}
\label{equ-pot}

We can derive from the system (\ref{Eqmedgen}) the equations satisfied by the potentials, using that
\begin{eqnarray*}
\mathbf{E} &=&\frac{1}{\varepsilon _{0}}\left( \mathbf{D}-\mathbf{P}\right)
=-\nabla \phi -\frac{\partial \mathbf{A}}{\partial t}, \\
\mathbf{B} &=&\frac{1}{\mu _{0}}\left( \mathbf{H}+\mathbf{M}\right) =\nabla
\times \mathbf{A}, \\
S &=&\mu _{0}\varepsilon _{0}\frac{\partial \phi }{\partial t}+\nabla \cdot 
\mathbf{A},
\end{eqnarray*}
resulting in 
\begin{subequations}
\label{eqspot}
\begin{eqnarray}
\mu _{0}\varepsilon _{0}\frac{\partial ^{2}\phi }{\partial t^{2}}-\nabla
^{2}\phi &=&\frac{1}{\varepsilon _{0}}\left( \varrho -\nabla \cdot \mathbf{P}
\right) , \\
\mu _{0}\varepsilon _{0}\frac{\partial ^{2}\mathbf{A}}{\partial t^{2}}
-\nabla ^{2}\mathbf{A} &=&\mu _{0}\left( \mathbf{j}+\nabla \times \mathbf{M}+
\frac{\partial \mathbf{P}}{\partial t}\right) ,
\end{eqnarray}
as one could have expected from the fact that the sources of $\mathbf{E}$ and $\mathbf{B}$ are the total (free plus molecular) charge and current densities.

From these equations, again with the constitutive relations $\mathbf{D}=\varepsilon \mathbf{E}$, $\mathbf{B}=\mu \mathbf{H}$, one readily obtains 
\end{subequations}
\begin{subequations}
\label{potentials}
\begin{eqnarray}
\mu _{0}\varepsilon _{0}\frac{\varepsilon _{0}}{\varepsilon }\frac{\partial
^{2}\phi }{\partial t^{2}}-\nabla ^{2}\phi &=&\frac{\rho }{\varepsilon }
+\left( 1-\frac{\varepsilon _{0}}{\varepsilon }\right) \frac{\partial }{
\partial t}\left( \nabla \cdot \mathbf{A}\right) , \\
\mu \varepsilon \frac{\partial ^{2}\mathbf{A}}{\partial t^{2}}-\nabla ^{2}
\mathbf{A} &=&\mu \mathbf{j}+\left( \frac{\mu }{\mu _{0}}-1\right) \nabla
\left( \nabla \cdot \mathbf{A}\right)  \notag \\
&&+\mu \varepsilon \left( \frac{\varepsilon _{0}}{\varepsilon }-1\right)
\nabla \left( \frac{\partial \phi }{\partial t}\right) .
\end{eqnarray}

If we apply these equations to a dielectric without losses, the amplitudes
of the Fourier modes satisfy 
\end{subequations}
\begin{eqnarray*}
\left( k^{2}-\mu _{0}\varepsilon _{0}\frac{\varepsilon _{0}}{\varepsilon }
\omega ^{2}\right) \phi _{0} &=&\left( 1-\frac{\varepsilon _{0}}{\varepsilon 
}\right) \omega \left( \mathbf{k}\cdot \mathbf{A}_{0}\right) , \\
\left( k^{2}-\mu \varepsilon \omega ^{2}\right) \mathbf{A}_{0} &=&\left( 1-\frac{\mu }{\mu _{0}}\right) \left( \mathbf{k}\cdot \mathbf{A}_{0}\right) 
\mathbf{k}+\mu \varepsilon \left( \frac{\varepsilon _{0}}{\varepsilon }
-1\right) \omega \mathbf{k}\phi _{0}.
\end{eqnarray*}

The transverse mode corresponds to an ordinary transverse electromagnetic wave in the gauge $\phi =0$, $\nabla \cdot \mathbf{A}=0$, propagating at the speed of light in the medium: $k^{2}=\mu \varepsilon \omega ^{2}$.

The longitudinal mode has the dispersion relation of a wave in vacuum: $k^{2}=\mu _{0}\varepsilon _{0}\omega ^{2}$. However, the remarkable point is
that the Fourier amplitude of the scalar: $S_{0}=i\left(\mathbf{k}\cdot 
\mathbf{A}_{0}-\mu _{0}\varepsilon _{0}\omega \phi _{0}\right) $, turns out to be zero for the corresponding longitudinal mode, and, consequently, also the longitudinal electric field is zero.

In the case of a conducting medium with the locally conserved current $\mathbf{j}=\sigma \mathbf{E}$, the equations for the Fourier amplitudes are 
\begin{eqnarray*}
\left( k^{2}-\mu _{0}\varepsilon _{0}\frac{\varepsilon _{0}}{\varepsilon }
\omega ^{2}\right) \phi _{0} &=&\left( 1-\frac{\varepsilon _{0}}{\varepsilon 
}\right) \omega \left( \mathbf{k}\cdot \mathbf{A}_{0}\right) +i\frac{\sigma 
}{\varepsilon }\left( \mathbf{k}\cdot \mathbf{A}_{0}-\frac{k^{2}}{\omega }
\phi _{0}\right) , \\
\left( k^{2}-\mu \varepsilon \omega ^{2}\right) \mathbf{A}_{0} &=&\left( 1-\frac{\mu }{\mu _{0}}\right) \left( \mathbf{k}\cdot \mathbf{A}_{0}\right) 
\mathbf{k}+\mu \varepsilon \left( \frac{\varepsilon _{0}}{\varepsilon }
-1\right) \omega \mathbf{k}\phi _{0} \\
&&+i\mu \sigma \left( \omega \mathbf{A}_{0}-\mathbf{k}\phi _{0}\right) .
\end{eqnarray*}

The corresponding transverse mode is of course (\ref{k2trans}), and the
longitudinal one the previously found paradoxical relation $k^{2}=\mu_{0}\varepsilon _{0}\omega ^{2}$, but again, as in the case of the
dielectric, one has $S_{0}=0$, and zero longitudinal electric field (and thus no dissipation, resolving the paradox).

In both, dielectric and conductor, the dispersion relation of the longitudinal mode, $k^{2}=\mu_{0}\varepsilon _{0}\omega ^{2}$, corresponds to a wave of pure potentials, with zero fields, in which the vector and scalar potentials are related by $\omega \mathbf{A}_{0}=\mathbf{k}\phi _{0}$, with corresponding $S_{0}=\mathbf{k}\cdot \mathbf{E}_{L0}=0$. We will denote this wave of pure potentials a ``gauge'' wave.

These results indicate that a scalar field wave cannot propagate in a material
medium in which local conservation of charge holds. If such a wave in vacuum converges on a medium it should be reflected, at most exciting in the medium a gauge wave propagating at the speed of light in vacuum. Ordinary transverse waves propagating in the material and in vacuum are in general also generated, as studied in some detail in the following Section.

\section{Reflection and transmission of waves at the vacuum-material interface}
\label{reflection}

\subsection{Incident scalar wave}\label{Inc_scalar}

We consider a scalar field wave in vacuum incident on a material surface at an
angle $\theta _{i}$, relative to the external normal. We consider the scalar
wave to have scalar and vector potentials $\phi _{i}$ and $\mathbf{A}_{i}$,
so that for the Fourier mode amplitudes we have 
\begin{equation*}
S_{0i}=-i\omega \phi _{0i}+i\mathbf{k}_{i}\cdot \mathbf{A}_{0i},
\end{equation*}
with corresponding longitudinal electric field $\mathbf{E}_{0i}=-i\mathbf{k}
_{i}\phi _{0i}+i\omega \mathbf{A}_{0i}$.

At the interface we must consider the presence of the incident wave ($i$),
reflected ones ($r$), and transmitted ones ($t$). As discussed above, the
longitudinal transmitted one must be a "gauge" wave, with
\begin{equation*}
\omega \mathbf{A}_{0t}=\mathbf{k}_{t}\phi _{0t}.
\end{equation*}

We must in general allow for the generation of usual transverse
radiation by the induced currents (free and/or of polarization). We know from the results in the previous Section that the transverse radiation in the medium has only a potential vector,
which for the Fourier amplitude of the transmitted wave we denote $\overline{
\mathbf{A}}_{0t}$, with wave-vector $\overline{\mathbf{k}}_{t}$, of angle $\overline{\theta }_{t}$. Without loss of generality, we consider the
reflected normal radiation also to be determined by only a transverse
potential vector \footnote{The most general transverse mode in vacuum can be considered as the superposition of a wave with only a transverse vector potential, and a gauge wave, with a longitudinal vector potential and a scalar potential
whose Fourier amplitudes are related by $\omega \mathbf{A}_{0}=\mathbf{k}\phi _{0}$. The same can be said of any other mode that has the dispersion relation $k^{2}=\mu _{0}\varepsilon _{0}\omega ^{2}$, since the addition of the gauge wave does not change the corresponding fields (this is the reason
of the name chosen for this wave).}, whose Fourier amplitude is denoted as $\overline{\mathbf{A}}_{0r}$, with wave-vector $\overline{\mathbf{k}}_{r}$, and angle $\overline{\theta }_{r}$. The reflected radiation also has a longitudinal component, in which a possible gauge component of the transverse reflected wave can be included.

Since the frequency $\omega $ is the same for all waves, and all
longitudinal, as well as the normal radiation in vacuum, have the same
dispersion relation: $k^{2}=\mu _{0}\varepsilon _{0}\omega ^{2}$, the
matching condition at the interface, which requires the equality of the
components parallel to the surface of all wave-vectors, indicates that the
angles between wave-vectors and surface normal are all equal for these four
waves; that is, $\theta _{r}=\overline{\theta }_{r}=\theta _{i}$, the usual
law of reflection, and $\theta _{t}=\theta _{i}$ (no refraction of the
longitudinal transmitted wave). On the other hand, the transverse
transmitted wave has the dispersion relation $k^{2}=\mu \varepsilon \omega
^{2}$, so that the matching condition is in this case
\begin{equation}
\sqrt{\mu \varepsilon }\sin \overline{\theta }_{t}=\sqrt{\mu _{0}\varepsilon
_{0}}\sin \theta _{i},  \label{snell}
\end{equation}
the usual Snell law.

The continuity of the potentials at the interface requires ($c$ is the speed of light in vacuum) 
\begin{subequations}
\label{BCpotentials}
\begin{eqnarray}
\mathbf{A}_{0i}+\mathbf{A}_{0r}+\overline{\mathbf{A}}_{0r} &=&\mathbf{A}%
_{0t}+\overline{\mathbf{A}}_{0t}, \\
\phi _{0i}+\phi _{0r} &=&\phi _{0t}=cA_{0t},
\end{eqnarray}
where $A_{0t}$ refers to the amplitude of the component of the vector potential, which is in the direction of its wave-vector: 
\end{subequations}
\begin{equation*}
\mathbf{A}_{0t}=A_{0t}\frac{\mathbf{k}_{t}}{\left\vert \mathbf{k}
_{t}\right\vert }.
\end{equation*}

Additional boundary conditions are the continuity of the tangential
component of the electric field, resulting from Eq. (\ref{faraday}), and
the continuity of the normal component of $\mathbf{B}$, which is identically
satisfied as it is zero at both sides. Finally, from Eqs. (\ref{Eqmedgen})
we have, for a dielectric, the continuity of the normal component of $\mathbf{D}$, and that of the tangential component of $\mathbf{H}$. For a
conductor, the same equations indicate that the normal component of $\mathbf{E}$ in vacuum must be equal to $\Sigma /\varepsilon _{0}$, with $\Sigma $
the surface charge density at the interface, and that the tangential
component of $\mathbf{B}$ in vacuum is equal to $\mu _{0}K_{T}$, with $K_{T} $ the tangential component of the surface current density at the interface.
Note that $\Sigma $ and $K_{T}$ are related by ($\nabla _{T}$ represents the
tangential derivative) 
\begin{equation*}
\frac{\partial \Sigma }{\partial t}+\nabla _{T}K_{T}=0.
\end{equation*}

Of course, in the case of a conductor there is no transmitted normal radiation ($\overline{\mathbf{A}}_{0t}=0$).

\begin{figure*}[h]
\includegraphics[width=10cm]{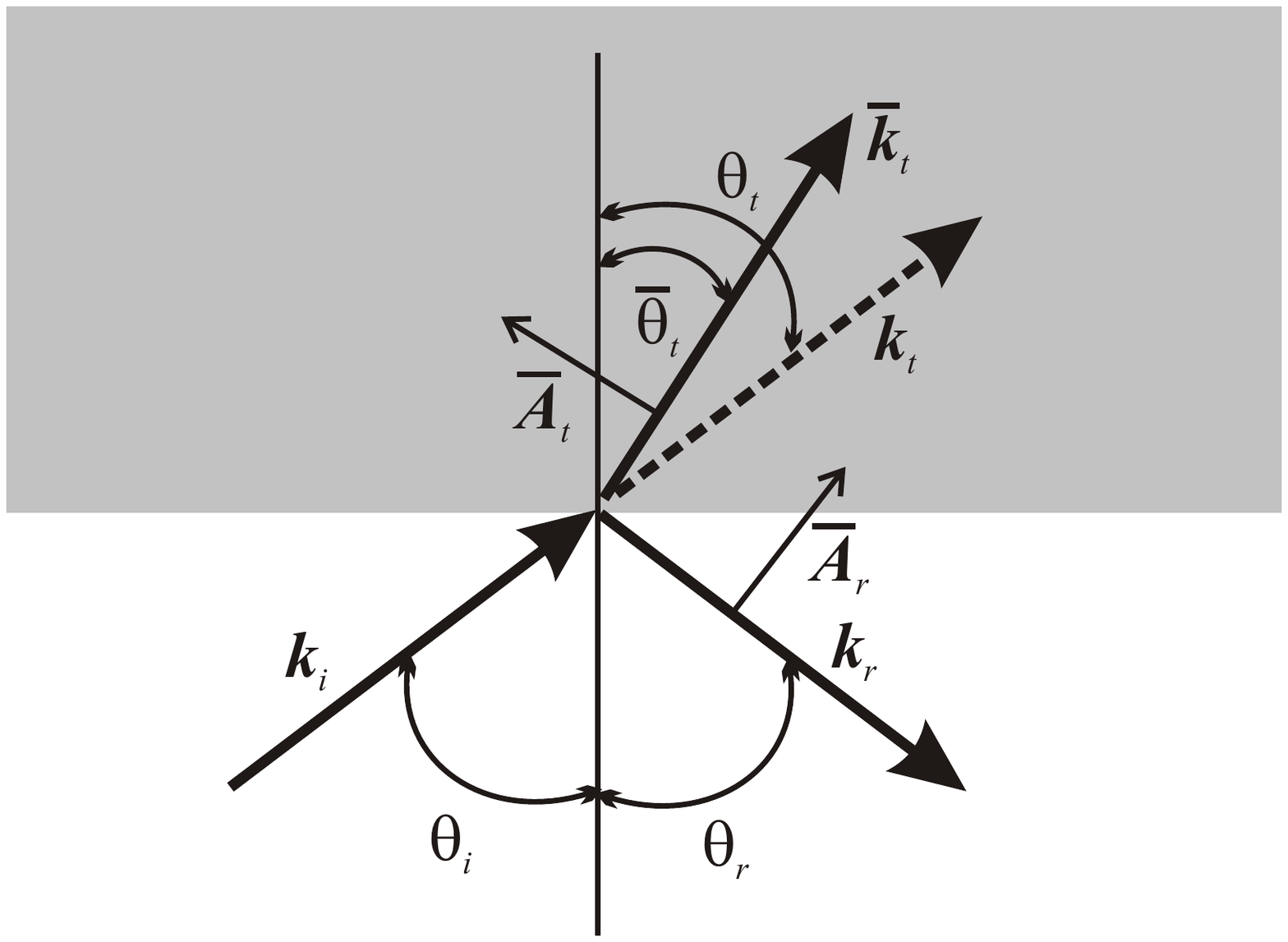}
\caption{Sketch of the incident ($i$), reflected ($r$), and transmitted ($t$) waves at the interface between vacuum (below) and material medium (above).
The overline denotes the transverse component parameters.\label{figbc}}
\end{figure*}

We refer to Fig. (\ref{figbc}) for the explicit expressions of the boundary
conditions. The vector potentials of the longitudinal modes have components
in the direction of their respective wave-vectors, and those of the
transverse modes have components in the directions explicitly denoted in the figure.

In the case of a dielectric medium the set of boundary conditions is then
written as:
\begin{eqnarray*}
A_{0i}\sin \theta _{i}+A_{0r}\sin \theta _{i}+\overline{A}_{0r}\cos \theta
_{i} &=&A_{0t}\sin \theta _{i}-\overline{A}_{0t}\cos \overline{\theta }_{t},
\\
A_{0i}\cos \theta _{i}-A_{0r}\cos \theta _{i}+\overline{A}_{0r}\sin \theta
_{i} &=&A_{0t}\cos \theta _{i}+\overline{A}_{0t}\sin \overline{\theta }_{t},
\\
\left( cA_{0i}-\phi _{0i}+cA_{0r}-\phi _{0r}\right) \sin \theta _{i}+c
\overline{A}_{0r}\cos \theta _{i} &=&-c\overline{A}_{0t}\cos \overline{
\theta }_{t}, \\
\left( -cA_{0i}+\phi _{0i}+cA_{0r}-\phi _{0r}\right) \cos \theta _{i}-c
\overline{A}_{0r}\sin \theta _{i} &=&-\frac{\varepsilon }{\varepsilon _{0}}c
\overline{A}_{0t}\sin \overline{\theta }_{t}, \\
\overline{A}_{0t}\sin \theta _{i} &=&\frac{\mu }{\mu _{0}}\overline{A}
_{0r}\sin \overline{\theta }_{t}, \\
\phi _{0i}+\phi _{0r} &=&\phi _{0t},
\end{eqnarray*}
which is to be completed with the Snell law (\ref{snell}).

For the case of a conductor we have $\overline{A}_{0t}=0$, and the boundary
conditions are
\begin{eqnarray*}
A_{0i}\sin \theta _{i}+A_{0r}\sin \theta _{i}+\overline{A}_{0r}\cos \theta
_{i} &=&A_{0t}\sin \theta _{i}, \\
A_{0i}\cos \theta _{i}-A_{0r}\cos \theta _{i}+\overline{A}_{0r}\sin \theta
_{i} &=&A_{0t}\cos \theta _{i}, \\
\left( cA_{0i}-\phi _{0i}+cA_{0r}-\phi _{0r}\right) \sin \theta _{i}+c
\overline{A}_{0r}\cos \theta _{i} &=&0, \\
ik\left( -cA_{0i}+\phi _{0i}+cA_{0r}-\phi _{0r}\right) \cos \theta _{i}-ikc
\overline{A}_{0r}\sin \theta _{i} &=&\frac{\Sigma _{0}}{\varepsilon _{0}}, \\
ik\overline{A}_{0r} &=&\mu _{0}K_{T0}, \\
\phi _{0i}+\phi _{0r} &=&\phi _{0t},
\end{eqnarray*}
to be completed by
\begin{equation*}
c\Sigma _{0}=K_{T0}\sin \theta _{i}.
\end{equation*}

The solution of these systems results in all Fourier amplitudes given in
terms of those of the incident wave, $\phi _{0i}$\ and $A_{0i}$.

For the case of a dielectric we give the expressions of the generated wave
amplitudes for the case of high-frequency waves, in particular optical
frequencies, so that the magnetization is negligible ($\mu \simeq \mu _{0}$), and the refraction index is thus given by $n=\sqrt{\varepsilon
/\varepsilon _{0}}$: 
\begin{subequations}
\label{pots_long_diel}
\begin{eqnarray}
\phi _{0r} &=&\left( \phi _{0i}-cA_{0i}\right) \frac{\left( n^{2}-1\right)
\sin ^{2}\theta _{i}}{n^{2}\cos ^{2}\theta _{i}+\cos \theta _{i}\sqrt{%
n^{2}-\sin ^{2}\theta _{i}}}, \\
\phi _{0t} &=&cA_{0t}=\phi _{0i}+\phi _{0r}, \\
A_{0r} &=&A_{0i}-\phi _{0i}/c+\phi _{0r}/c, \\
\overline{A}_{0r} &=&\left( \phi _{0i}/c-A_{0i}\right) \frac{2n^{2}\sin
\theta _{i}}{n^{2}\cos \theta _{i}+\sqrt{n^{2}-\sin ^{2}\theta _{i}}}, \\
\overline{A}_{0t} &=&\frac{\overline{A}_{0r}}{n},
\end{eqnarray}

In the case of a conductor we have 
\end{subequations}
\begin{subequations}
\label{pots_long_cond}
\begin{eqnarray}
\phi _{0r} &=&\left( \phi _{0i}-cA_{0i}\right) \tan ^{2}\theta _{i}, \\
\phi _{0t} &=&cA_{0t}=\phi _{0i}+\phi _{0r}, \\
A_{0r} &=&\left( A_{0i}-\phi _{0i}/c+\phi _{0r}/c\right) , \\
\overline{A}_{0r} &=&2\left( \phi _{0i}/c-A_{0i}\right) \tan \theta _{i}.
\end{eqnarray}

The extra source $I$ involved in the reflection and transmission of the longitudinal wave can be easily determined from the wave equation of the scalar field 
\end{subequations}
\begin{equation*}
\mu _{0}\varepsilon _{0}\frac{\partial ^{2}S}{\partial t^{2}}-\nabla
^{2}S=\mu _{0}I,
\end{equation*}
integrated in the conventional pill-box to give for the Fourier amplitudes ($k=\left\vert \mathbf{k}_{i}\right\vert =\left\vert \mathbf{k}_{r}\right\vert 
$)
\begin{eqnarray*}
\mu _{0}\Gamma _{0} &=&ik\cos \theta _{i}\left( S_{0i}-S_{0r}\right) \\
&=&-k^{2}\sin \theta _{i}\left( A_{0i}-\frac{\phi _{0i}}{c}-A_{0r}+\frac{
\phi _{0r}}{c}\right) ,
\end{eqnarray*}
where the Fourier amplitude $\Gamma _{0}$ is related to that of the extra
source by $\Gamma _{0}=I_{0}\delta $, with $\delta $ the depth of the
region occupied by the non-conserved current. The remarkable result is that
for both, dielectric and conductor, the required extra source is zero. This
means that the ordinary sources in the medium are sufficient to generate the
mode conversion of the incident scalar wave, leading to no propagation of a
scalar wave in the bulk of the medium. The only possible longitudinal mode transmitted is a gauge wave, as shown by the second equation in each of relations (\ref{pots_long_diel}) and (\ref{pots_long_cond}).

\subsection{Incident transverse field wave}

For the usual Maxwell transverse field wave, the fields satisfy the
Fresnel relations for reflected and transmitted waves, in the case of a
dielectric, and the reflected field laws in case of a conductor. In Aharonov-Bohm electrodynamics we
must of course consider also the behavior of the corresponding potentials. For
this, we must take into account that the potentials of a transverse field
wave correspond to a superposition of a pure transverse vector potential and
a gauge wave. We thus denote the incident wave potentials Fourier amplitudes
by $\overline{A}_{0i}$, for the transverse vector potential, and $\phi
_{0i}=cA_{0i}$ for the potentials of the gauge wave. Correspondingly, the
reflected transverse field wave have a transverse vector potential amplitude 
$\overline{A}_{0r}$, and gauge part with potentials $\phi _{0r}=cA_{0r}$. In
case of a dielectric there is also a transmitted transverse field wave, with
transverse vector potential $\overline{A}_{0t}$, and gauge component with
potentials $\phi _{0t}=cA_{0t}$.

\begin{figure*}[h]
\includegraphics[width=10cm]{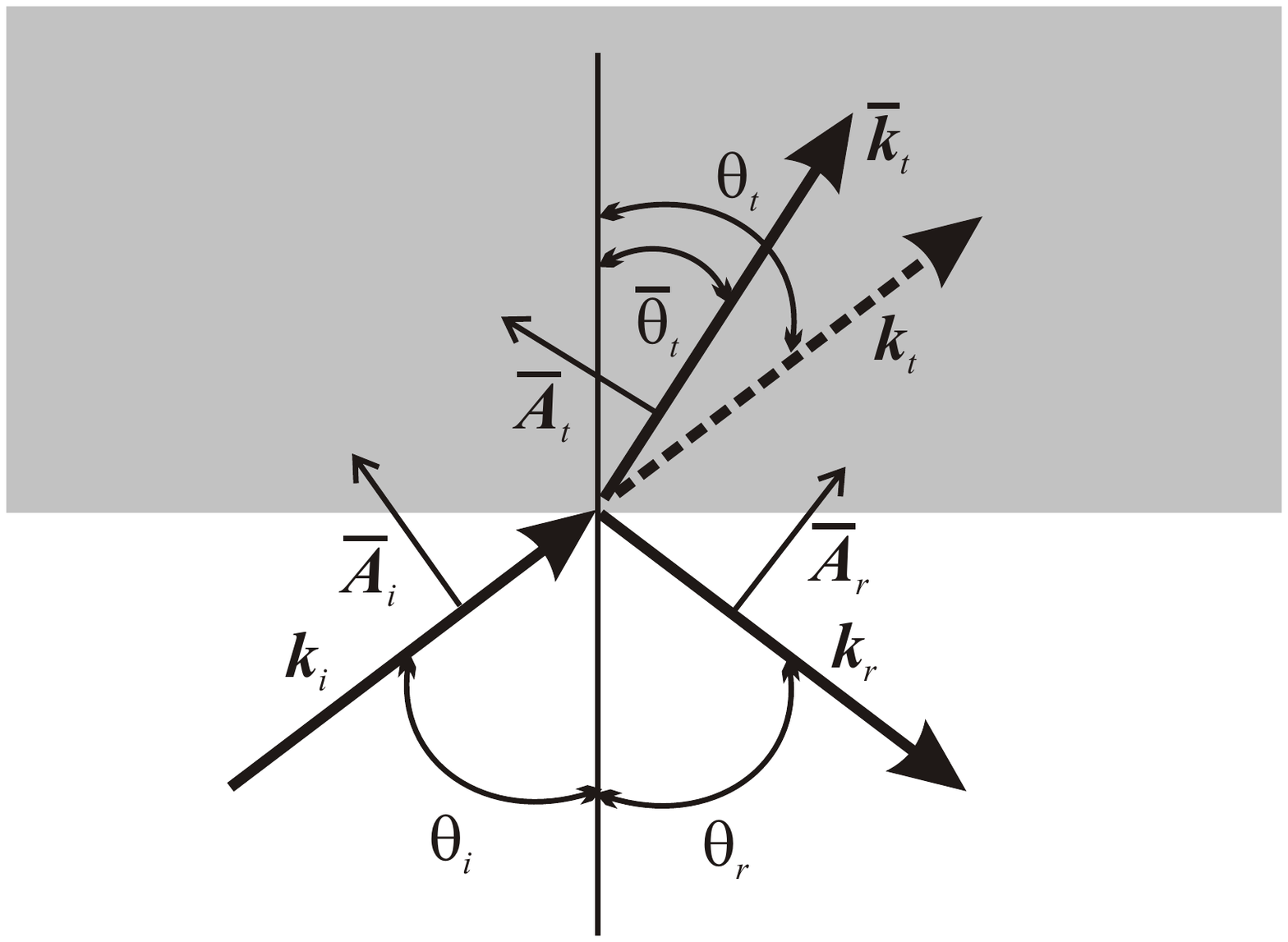}
\caption{Convention used for the case of a dielectric with a transverse incident wave
whose polarization corresponds to the magnetic field parallel to the interface (normal to the plane of the figure).\label{figB-parallel}}
\end{figure*}

The polarization of the incident wave must also be taken into account. We
thus divide the problem into one with a polarization with magnetic field
parallel to the interface, and another with electric field parallel to the
interface. See Fig. (\ref{figB-parallel}) for the conventions used in the
case with $\mathbf{B}$ parallel to the interface. As before, the transverse
component of the vector potentials are in the directions indicated in the
figure, while their longitudinal components are in the direction of their
corresponding wave-vectors.

We consider first the case of a dielectric.

For a polarization with the magnetic field of the wave parallel to the
interface, the conditions of continuity of potentials and of the fields, as
determined by Eqs. (\ref{Eqmedgen}), are then written as%
\begin{eqnarray*}
-\overline{A}_{0i}\cos \theta _{i}+A_{0i}\sin \theta _{i}+A_{0r}\sin \theta
_{i}+\overline{A}_{0r}\cos \theta _{i} &=&A_{0t}\sin \theta _{i}-\overline{A}
_{0t}\cos \overline{\theta }_{t}, \\
\overline{A}_{0i}\sin \theta _{i}+A_{0i}\cos \theta _{i}-A_{0r}\cos \theta
_{i}+\overline{A}_{0r}\sin \theta _{i} &=&A_{0t}\cos \theta _{i}+\overline{A}
_{0t}\sin \overline{\theta }_{t}, \\
\overline{A}_{0i}\cos \theta _{i}-\overline{A}_{0r}\cos \theta _{i} &=&%
\overline{A}_{0t}\cos \overline{\theta }_{t}, \\
\mu \left( \overline{A}_{0i}+\overline{A}_{0r}\right) \sin \overline{\theta }
_{t} &=&\mu _{0}\overline{A}_{0t}\sin \theta _{i},
\end{eqnarray*}
which, for the case of high-frequency waves ($\mu \simeq \mu _{0}$, $n=\sqrt{
\varepsilon /\varepsilon _{0}}$), result in :%
\begin{eqnarray*}
\overline{A}_{0r} &=&\overline{A}_{0i}\frac{n^{2}\cos \theta _{i}-\sqrt{
n^{2}-\sin ^{2}\theta _{i}}}{n^{2}\cos \theta _{i}+\sqrt{n^{2}-\sin
^{2}\theta _{i}}}, \\
\overline{A}_{0t} &=&\overline{A}_{0i}\frac{2n\cos \theta _{i}}{n^{2}\cos
\theta _{i}+\sqrt{n^{2}-\sin ^{2}\theta _{i}}}, \\
A_{0r} &=&\frac{\phi _{0r}}{c}=\overline{A}_{0i}\frac{\left( n^{2}-1\right)
\sin \theta _{i}}{n^{2}\cos \theta _{i}+\sqrt{n^{2}-\sin ^{2}\theta _{i}}},
\\
A_{0t} &=&\frac{\phi _{0t}}{c}=A_{0i}+A_{0r}.
\end{eqnarray*}

\begin{figure*}[h]
\includegraphics[width=10cm]{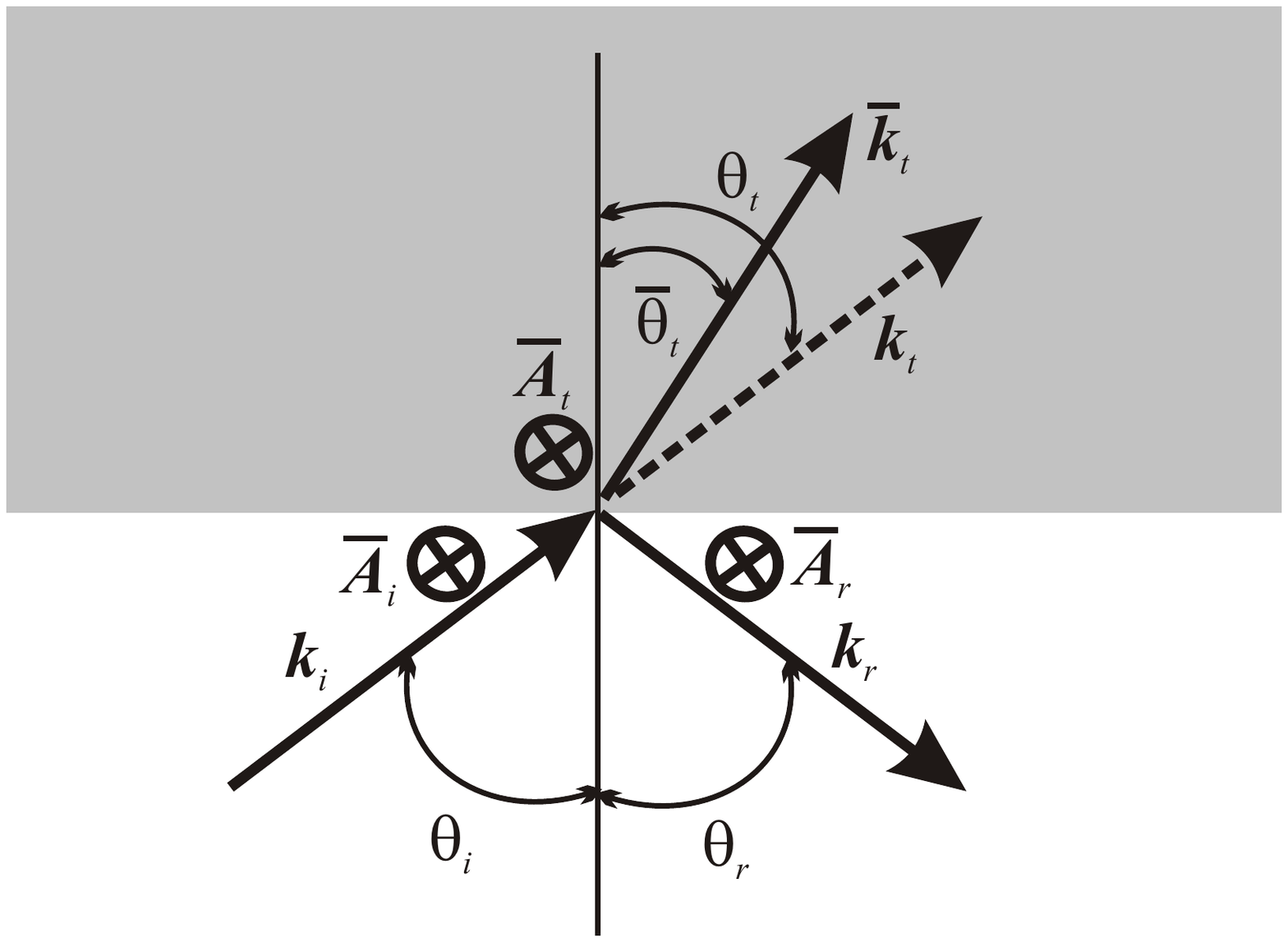}
\caption{Convention used for the case of a dielectric
with a transverse incident wave whose polarization corresponds to the electric field parallel to the interface (normal to the plane of the figure).\label{fig-Eparallel}}
\end{figure*}

For a polarization with the electric fields parallel to the interface, we refer to Fig. (\ref{fig-Eparallel}) for the conventions used in this case,

\begin{eqnarray*}
A_{0i}+A_{0r} &=&A_{0t}, \\
A_{0i}-A_{0r} &=&A_{0t}, \\
\overline{A}_{0i}+\overline{A}_{0r} &=&\overline{A}_{0t}, \\
\mu _{0}\overline{A}_{0t}\sin \theta _{i}\cos \overline{\theta }_{t} &=&\mu
\left( \overline{A}_{0i}-\overline{A}_{0r}\right) \cos \theta _{i}\sin 
\overline{\theta }_{t},
\end{eqnarray*}
resulting in
\begin{eqnarray*}
A_{0t} &=&\frac{\phi _{0t}}{c}=A_{0i}, \\
A_{0r} &=&\frac{\phi _{0r}}{c}=0, \\
\overline{A}_{0r} &=&\overline{A}_{0i}\frac{\cos \theta _{i}-\sqrt{%
n^{2}-\sin ^{2}\theta _{i}}}{\cos \theta _{i}+\sqrt{n^{2}-\sin ^{2}\theta
_{i}}}, \\
\overline{A}_{0t} &=&\overline{A}_{0i}\frac{2\cos \theta _{i}}{\cos \theta
_{i}+\sqrt{n^{2}-\sin ^{2}\theta _{i}}}.
\end{eqnarray*}

For a conductor, in the case of polarization with the magnetic field
parallel to the interface the continuity conditions give
\begin{eqnarray*}
-\overline{A}_{0i}\cos \theta _{i}+A_{0i}\sin \theta _{i}+A_{0r}\sin \theta
_{i}+\overline{A}_{0r}\cos \theta _{i} &=&A_{0t}\sin \theta _{i}, \\
\overline{A}_{0i}\sin \theta _{i}+A_{0i}\cos \theta _{i}-A_{0r}\cos \theta
_{i}+\overline{A}_{0r}\sin \theta _{i} &=&A_{0t}\cos \theta _{i}, \\
\overline{A}_{0i}-\overline{A}_{0r} &=&0,
\end{eqnarray*}
with the result
\begin{eqnarray*}
A_{0t} &=&\frac{\phi _{0t}}{c}=A_{0i}+\overline{A}_{0i}\tan \theta _{i}, \\
A_{0r} &=&\frac{\phi _{0r}}{c}=\overline{A}_{0i}\tan \theta _{i}, \\
\overline{A}_{0r} &=&\overline{A}_{0i}.
\end{eqnarray*}

If the polarization of the wave is with the electric fields parallel to the
interface the boundary conditions are
\begin{eqnarray*}
A_{0i}+A_{0r} &=&A_{0t}, \\
A_{0i}-A_{0r} &=&A_{0t}, \\
\overline{A}_{0i}+\overline{A}_{0r} &=&0,
\end{eqnarray*}
with the result 
\begin{eqnarray*}
A_{0t} &=&\frac{\phi _{0t}}{c}=A_{0i}, \\
A_{0r} &=&\frac{\phi _{0r}}{c}=0, \\
\overline{A}_{0r} &=&-\overline{A}_{0i}.
\end{eqnarray*}

The transverse vector potential components in the relations obtained in this Subsection correspond to fields that satisfy the above mentioned known laws for the Maxwell transverse waves: Fresnel relations for a dielectric medium, and reflection laws for a conductor. The most remarkable result is the generation of transmitted and reflected gauge waves when a Maxwell transverse wave interacts with a dielectric or a conductor.

\section{Conclusions}\label{conclusions}

In this work we have analyzed with some detail the interaction of different waves, possible in the extended electrodynamics of Aharonov-Bohm, with material media in which the local conservation of current (free and/or molecular) holds. A particularly interesting result is the impossibility of the transmission through those media of the longitudinal modes that, according to that theory, are possible in vacuum. Only the so-called ``gauge waves" propagate in the medium, as shown in Subsect. \ref{Inc_scalar}.

It is important to mention that the gauge wave is not just a pictorial representation of the reduced gauge freedom of the theory, as could be used, for instance, to represent the addition of such a wave to another type of wave in order to effect an allowed change of its gauge, without change of the wave fields (electric, magnetic and scalar). According to the theory, a gauge wave can be generated and detected by itself. In other words, the theory allows pure gauge waves, without another accompanying type of wave, as physical entities that can in principle be generated and detected.

As can be seen from Eqs. (\ref{pots_long_diel}) and (\ref{pots_long_cond}), when such a pure gauge wave reaches a medium (that is, the only incident potentials satisfy $\phi_{0i}=cA_{0i}$) it is fully transmitted without generation of additional waves. This is of course consistent with its not interacting with media in which local conservation of charge holds.

Concerning the generation of a gauge wave, from the results for a transverse wave incident in a medium we see that a normal transverse wave can generate a gauge wave when it interacts with that medium. Consider for instance the potentials of a normal transverse wave generated by an oscillating elementary dipole $\mathbf{p}\left( t\right) $ (assumed at the origin of coordinates) 
\begin{eqnarray*}
\phi \left( \mathbf{x},t\right) &=&\frac{\mu _{0}c}{4\pi r}\overset{.}{\mathbf{p}}\left( t-r/c\right) \cdot \mathbf{n}, \\
\mathbf{A}\left( \mathbf{x},t\right) &=&\frac{\mu _{0}}{4\pi r}\overset{.}{\mathbf{p}}\left( t-r/c\right) ,
\end{eqnarray*}
where $r=\left\vert \mathbf{x}\right\vert $, and $\mathbf{n}=\mathbf{x}/r$.
We see that for a given propagation direction $\mathbf{n}$, there is a field-transverse wave consisting in the superposition of a transverse vector potential ($\mathbf{I}$ is the identity matrix) 
\begin{equation*}
\mathbf{A}_{\perp }\left( \mathbf{x},t\right) =\frac{\mu _{0}}{4\pi r} 
\overset{.}{\mathbf{p}}\left( t-r/c\right) \cdot \left( \mathbf{I}-\mathbf{nn}\right) ,
\end{equation*}
and a gauge wave with
\begin{equation*}
A_{\parallel }\left( \mathbf{x},t\right) =\frac{\phi \left( \mathbf{x},t\right) }{c}=\frac{\mu _{0}}{4\pi r}\overset{.}{\mathbf{p}}\left(
t-r/c\right) \cdot \mathbf{n}.
\end{equation*}

In particular, we see that in the direction of the dipole a pure gauge wave is emitted, while at right angles what is emitted is a pure transverse vector potential. For intermediate angles a mixture of components is involved. Interaction of the waves emitted at these intermediate angles with a medium would thus generate gauge waves, additional to those emitted in the direction of the dipole.

It is important to recall \cite{minotti2021quantum} that what one denotes as the gauge freedom of Aharonov-Bohm electrodynamics has a different meaning from that in Maxwell's. In the latter theory, even if all electromagnetic sources are known, one is still free to select different gauges. In the former theory, however, if all sources are known (in the example above the only source is an elementary dipole) the gauge is completely determined. The reduced gauge freedom allows only to include ``incoming" potentials (as boundary conditions) generated by sources other than those in the region considered. In this way, the gauge wave components determined in the example of the dipole are unique, a consequence of the theory that lends further support to the ``reality" of the gauge wave.

As for the detection of a pure gauge wave, from the results in \cite{minotti2021quantum} one concludes that this type of wave can only interact with media in which the extra source $I$ is not zero, resulting in a power per unit volume $w=-I\phi $, exchanged between the wave and the medium supporting the extra current, that could be detected as changes in currents and voltages in the circuit. Since collective quantum effects are likely required for this \cite{minotti2022electromagnetic}, we call such a circuit, for brevity sake, a ``quantum circuit".

Since specific examples of a quantum circuit require detailed models of media in which charge is not locally conserved, an area in which much theoretical and experimental work is needed, we postpone the discussion of the matter to a future work.

A final consideration is that the gauge wave would be a very desirable means of communication, since it could propagate through any conventional medium without being dissipated or deviated. A modulated (in amplitude or frequency) gauge wave could in principle be detected by a properly designed quantum circuit.

\bibliographystyle{ieeetr}
\bibliography{ems-ABED}

\end{document}